\documentclass[12pt, fleqn]{article}
\usepackage[cp1251]{inputenc}
\usepackage{latexsym,amsfonts,amssymb}
\usepackage{graphicx}

\usepackage{amsbsy}
\usepackage{amsmath}
\usepackage{epsf}
\usepackage{cite}

\newcommand{\plussupset}{%
  \mathrel{%
    \ooalign{%
      $\supset$\cr
      \hidewidth\kern-0.1em\raisebox{0.09ex}{$\scriptstyle+$}\hidewidth\cr
    }%
  }%
}

\newtheorem{theorem}{Theorem}

\newtheorem{remark}{Remark}
\newcommand{\bt}{\begin{theo}}
\newcommand{\et}{\end{theo}}
\newcommand{\bd}{\begin{displaymath}}
\newcommand{\ed}{\end{displaymath}}

\newcommand{\lf}{\left}
\newcommand{\rg}{\right}

\newcommand{\be} {\begin{equation}}
\newcommand{\ee} {\end{equation}}
\newcommand{\ba} {\begin{array}{l}}
\newcommand{\ea} {\end{array}}
\newcommand{\bea}{\begin{eqnarray}}
\newcommand{\eea} {\end{eqnarray}}

\newcommand{\p} {\partial}

\begin{document}

\begin{center}
 {\Large \bf
Analysis of a  mathematical model for fluid  transport  in poroelastic materials in 2D space }

\medskip

{\bf Roman Cherniha $^{a}$\footnote{\small  Corresponding author.
E-mail: r.m.cherniha@gmail.com},
  Vasyl' Davydovych$^{b}$, Stachowska-Pietka$^{c}$   and Jacek~Waniewski$^{c}$}

$^{a}$  \quad National University of Kyiv-Mohyla Academy, 2,
Skovoroda Street, Kyiv  04070, Ukraine.

$^{b}$ \quad Institute of Mathematics,  National Academy of Sciences
of Ukraine, \\
 3, Tereshchenkivs'ka Street, Kyiv 01004, Ukraine.\\

$^{c}$ \quad Institute of Biocybernetics and Biomedical Engineering, PAS,
    Ks. Trojdena 4, 02 796 Warsaw, Poland

\end{center}

\begin{abstract}
Mathematical models that are based on  the poroelastic theory are
applied in biology, medicine, and biomedical engineering to describe a wide range of processes. In this work,  a
mathematical model for poroelastic materials (PEM) with
variable volumes is developed in a multidimensional case. Governing
equations of the model are constructed using  continuity
equations, which reflect the well-known physical laws. The
deformation vector is specified using the Terzaghi effective stress
tensor. In the  two-dimensional space case, the    model is studied
by analytical methods. Using the classical Lie method, it is proven
that the relevant nonlinear system of the $(1+2)$-dimensional
governing equations admits highly nontrivial Lie symmetries which lead
to an infinite-dimensional Lie algebra. The radially-symmetric case
is studied in details. It is shown how correct boundary conditions
in the case of PEM in the form of a ring and an annulus are
constructed. As a result, boundary-value  problems  with a moving
boundary describing the ring (annulus) deformation  are constructed.
 The relevant
nonlinear boundary-value  problems  are analytically solved in the
stationary case. In particular, the analytical formulae for unknown
deformations and an unknown radius of the annulus  are derived.
Moreover, illustrative  plots   for parameters, which are typical for the
tumor tissue deformation, are presented.

\end{abstract}

\section{Introduction}  \label{sec-1}

The modern  poroelastic theory  is well presented in literature  \cite{
taber-04,loret-simoes-17, deto-93,coussy-1995,coussy-2010}.
Mathematical models that were created on the basis of this theory
are widely  applied in geology to describe the  penetration of
water or oil across soil or cracked rock  (see, e.g.,  \cite{lacis-17,siddique-17,guer-2021}).  Moreover, there  are many
applications of the  poroelastic theory in biology, medicine,
and biomedical engineering to describe  a wide range of
processes (see, e.g., \cite{ loret-simoes-17, speciale-2008,
speciale-2009,travas-2014, mascheroni-2018,
sowin-2021,ch-wa-2020, ch-wa-2022, ch-st-wa-24}).

First, we point out that the poroelastic theory uses several notions
(displacement, stress tensor, strain tensor, etc.) from the
elasticity theory applied to the materials of continuum mechanics
(see, e.g., \cite{samaniego-et-al-2020, soldatos-2025} and papers
cited therein). However, in contrast to the elasticity theory,  the
poroelastic theory considers the system formed by a solid, elastic
matrix with pores that can be penetrated by a fluid and dissolved
solutes (e.g., water with high concentration of salts). The
poroelastic material (PEM) is considered as the superposition of two
continious media: the matrix (skeleton) and the system of  pores
saturated by a fluid.  The deformation of the system under the fluid
pressure is described by a deformation vector, and the dynamics of
the deformation under the forces needs, in general, to be described
by second order tensors.
 The relationship between stress and strain
is usually assumed to be linear. The flux of fluid depends on the
hydrostatic pressure and fluid (solute, oil)  gradients (called
osmotic pressure).
Additionally, the   diffusive and convectivex
transport mechanisms should be taking into account.

 The general three-dimensional theory describing  fluid
  transport in PEM is
very complex
 because relevant
mathematical models involve 3D nonlinear  partial differential
equations (PDEs). Moreover, if one takes  time-dependence into account,
then 4D PDEs should be used. As a result, its reduction to two- or
one-dimensional version is often discussed
 \cite{ch-wa-2020,ch-wa-2022,ch-st-wa-24,ch-da-vo-24,daria-et-al-2001,li-schanz-11}.
 In particular,  the relevant mathematical models  can be
analytically solved in some cases  (at least under some assumptions). In
contrast to the above cited papers, here, a multidimensional model is
developed and  analytical results are derived in the 2D space case.
It  should be stressed  that analytical results, especially  exact
solutions, do  help to understand the role of different parameters
and are useful to estimate  the accuracy of numerical simulations.

Nowadays,  the most powerful  methods to construct the  exact
solutions to nonlinear PDEs are  symmetry-based methods, in
particular,  the Lie method and its generalizations. There are
thousands of published papers devoted to the application of
symmetry-based methods to PDEs; therefore, we list only some recent
monographs, such as \cite{bl-anco-10, arrigo15, ch-se-pl-2018}. It
should be pointed out that the authors typically examine scalar PDEs
because the search for symmetries of {\it systems of nonlinear PDEs}
is a much complicated problem. In fact, essential technical
difficulties occur if one intends to identify symmetries and
construct exact solutions for systems of PDEs. A typical example is
the very recent study \cite{mandal-2025}, in which the authors
considered  a three-component system arising in fluid dynamics for
the very beginning. However, in order to identify symmetries, they
simplified  the system to  a single fourth-order PDE and were only
able to obtain  particular results about symmetries of the PDE in
question \cite{ch-arxiv-2025}.  There are some rigorous studies
devoted to the application of symmetry-based methods to systems of
nonlinear  PDEs, in particular, the papers devoted to {\it
multicomponent systems of PDEs} (notably, the model developed below
is such a system).
   Taking  the above observation into account, we  refer  the reader to the recent works
 \cite{ch-dav-2019,nadjaf-2021,ch-dav-2022,pal-2023,shag-2023,oliveri-25},
which are   devoted to applications of symmetry-based
 methods  to the nonlinear  multicomponent systems of PDEs.



This work is organized as follows. In Section~\ref{sec:2}, a
mathematical model for PEM with
variable volumes is developed in a multidimensional case.  Governing
equations of the model are constructed using the continuity
equations, which reflect the well-known physical laws. The
deformation vector is specified using  the Terzaghi effective stress
tensor (see, e.g.,
\cite{loret-simoes-17,coussy-2010,terzaghi-1936}). In
Section~\ref{sec:3}, the model is studied in the 2D space
approximation. Using the classical Lie method, we  prove that the
governing equations, which form a six-component system of nonlinear
$(1+2)$-dimensional PDEs, admit highly nontrivial Lie symmetries, which lead  to infinite-dimensional Lie algebra. In Section~\ref{sec:4},
the radially-symmetric case is studied. It is shown how  correct
boundary conditions are constructed in the case of PEM in the form
of a ring that is shrinking. As a result, the boundary-value problem
(BVP) with a moving boundary describing the ring (annulus)
deformation  is constructed. In Section~\ref{sec:5}, the nonlinear
BVP obtained in the previous section is analytically solved in the
stationary case. It is shown how  deformation  and an unknown radius
of PEM after shrinking can  be exactly calculated by relevant
formulae. Illustrative plots  are presented for the parameters, which
are  typical   for the   tumor tissue deformation. Finally, we
discuss the results obtained and present some conclusions  in the
last section.

\section{ A model for fluid and solute transport in PEM} \label{sec:2}

The mathematical model for PEM with variable volumes is developed under the following assumptions:
\begin{enumerate}
\item[\rm{(1)}] PEM consists of pores and matrix;
\item[\rm{(2)}] no internal sources/sinks;
\item[\rm{(3)}] deformation is described by a linear stress tensor;
\item[\rm{(4)}] incompressible fluid;
\item[\rm{(5)}] isothermal conditions for  transport in PEM.
\end{enumerate}

 The governing equations of the model consist of continuity equations.

 Let us consider  an infinitesimal  element $dV$ of PEM without division
 on two phases. The volume balance is as follows  (see  Chapter 10 in \cite{loret-simoes-17} and \cite{ch-st-wa-24}):
    \be \label{2-1*} \frac{1}{{dV}}\frac{{\partial (dV)}}{{\partial t}} =
    \frac{{\partial e}}{{\partial t}} = \frac{\partial\nabla \cdot \bar{u}  }{\partial t}= - \nabla \cdot  {\bar j_V}.\ee
Here, the deformation vector $u$ is defined as follows: \be \label{2-2*} u(
\bar x,t; \bar X) = \bar x( t ) - \bar X, \,\bar X = \bar x\left( 0
\right), \quad e= \nabla \cdot \bar{u}, \ee
where
  ${\bar j_V}$ is the volumetric flux across the PEM  (will be determined later),
  $\bar X=(X_1,\dots,X_n)$ is the initial position within PEM, and
  $\bar x(t)=(x_1(t),\dots,x_n(t))$ is the position at time $t$ (i.e., ~after displacement (deformation)).

Consider the mass $\rho dV$ of the infinitesimal element $dV$.
The mass conservation law gives the following continuity equation:
\[\frac{1}{{dV}}\frac{{\partial \left( {\rho dV} \right)}}{{\partial t}} =
\frac{{\partial \rho }}{{\partial t}} + \rho \frac{{\partial
e}}{{\partial t}} = - \nabla \cdot{\bar j_\rho }.\]

Thus, \be \label{2-3*} \frac{{\partial \rho }}{{\partial t}} = -
\nabla\cdot {\bar j_\rho } + \rho \nabla\cdot {\bar j_V},\ee  where
$\bar j_\rho$ is the flux of mass
 through the~PEM (will be determined later).

 On the other hand, one may consider the infinitesimal  element $dV$
 as a PEM element consisting of
  the fluid phase $F$
 with~the fractional volume (porosity) ${\theta _F}$ and the density $\rho _F$, and the solid (matrix) phase  $M$ with~the fractional volume ${\theta _M}$ and the density $\rho _M$.
  Obviously,
 \be \label{2-8} \rho = {\rho _F}{\theta _F} + {\rho _M}{\theta _M}, \quad  {\theta _F} + {\theta _M}=1, \ee
with $\rho_F=\rho_F^0=const$ (incompressibility).

 The volume balance equations are
\[\frac{1}{{dV}}\frac{{\partial \left( {{\theta _F}dV}
\right)}}{{\partial t}} = \frac{{\partial {\theta _F}}}{{\partial
t}} + {\theta _F}\frac{{\partial e}}{{\partial t}} = - \nabla\cdot
\bar j_{VF}, \ \frac{1}{{dV}}\frac{{\partial \left( {{\theta
_M}dV} \right)}}{{\partial t}} = \frac{{\partial {\theta
_M}}}{{\partial t}} + {\theta _M}\frac{{\partial e}}{{\partial t}} =
- \nabla \cdot\bar j_{VM},\]  so that \be \label{2-4}
\frac{{\partial {\theta _F}}}{{\partial t}} = - \nabla \cdot {\bar
j_{VF}} + {\theta _F}\nabla \cdot {\bar j_V},\ee \be \label{2-5}
\frac{{\partial {\theta _M}}}{{\partial t}} = - \nabla \cdot {\bar
j_{VM}} + {\theta _M} \nabla \cdot {\bar j_V},\ee where
 $\bar j_{VF}$  is the volumetric flux in the fluid phase (will be determined
 later), and
 $\bar j_{VM}$  is  the volumetric flux in the matrix phase (will be determined later).

The mass conservation law gives the following equations for each
phase: \be \label{2-6} \frac{1}{{dV}}\frac{{\partial \left( {{\rho
_F}{\theta _F}dV} \right)}}{{\partial t}} = \frac{{\partial \left(
{{\rho _F}{\theta _F}} \right)}}{{\partial t}} + {\rho _F}{\theta
_F}\frac{{\partial e}}{{\partial t}} = - \nabla \cdot \left( {{\rho
_F}{{\bar j}_{VF}}} \right),\ee \be \label{2-7}
\frac{1}{{dV}}\frac{{\partial \left( {{\rho _M}{\theta _ M}dV}
\right)}}{{\partial t}} = \frac{{\partial \left( {{\rho _M}{\theta
_M}} \right)}}{{\partial t}} + {\rho _M}{\theta _M}\frac{{\partial
e}}{{\partial t}} = - \nabla \cdot \left( {{\rho _M}{{\bar j}_{VM}}}
\right).\ee

Since we assume that the fluid phase is  a solution (not  pure water)
with the concentration $c(t,x)$ (e.g., glucose molecules), the
corresponding continuity equation can be written as follows:
    \be \label{2-9} \frac{1}{{dV}}\frac{{\partial \left( {c{\theta _F}dV} \right)}}{{\partial t}} = \frac{{\partial \left( {c{\theta _F}} \right)}}{{\partial t}} + c{\theta _F}\frac{{\partial e}}{{\partial t}} = - \nabla \cdot {\bar j_S},\ee
    where  $\bar j_S $ is the solute flux across the PEM.

Because the  PEM is deformed with time,  the~dynamics of the PEM
element of the mass $\rho dV$ with the velocity $ \frac{\partial
u}{\partial t} $  can be  described by the Newton law as follows:
    \be \label{2-10} \frac{1}{{dV}}\frac{{\partial \left( {\rho \frac{{\partial u}}{{\partial t}}dV} \right)}}{{\partial t}} =
     \rho \frac{{{\partial ^2}u}}{{\partial {t^2}}} + \frac{{\partial u}}{{\partial t}}\left( {\frac{{\partial \rho }}{{\partial t}} +
     \rho \frac{{\partial e}}{{\partial t}}} \right) = \nabla \cdot {\tilde \tau},\ee
    where ${\tilde \tau }$ is an  effective stress tensor.
In the general case, ${\tilde \tau }$  should be the Cauchy
effective stress tensor with possible nonlinear terms  w.r.t. the
dilatation $e=\nabla \cdot {\bar u}$ \cite{coussy-2010}.
 However, if  we restrict ourselves to
 the linear poroelastic theory for isotropic materials, then it is
the Terzaghi effective stress tensor (see, e.g.,
\cite{loret-simoes-17,coussy-2010,terzaghi-1936})
 \be \label{2-11}{\tilde \tau _t} = - p^*I + E\varepsilon= - p^*I + \lambda Tr \varepsilon I
 +2\mu \varepsilon.\ee
 Here,  $p^*= p - \sigma_1c$ is the effective pressure  given by the difference between hydrostatic
 pressure and osmotic pressure,  $\sigma_1$ is a constant parameter (see below for more details), $\lambda$ and $\mu$
   are the Lame coefficients, and
 $\varepsilon$ represents the strain tensor in the form of the symmetric  matrix $n\times
 n$; in particular, for $n=2$,
  \[
 \varepsilon=\frac{1}{2}\left(\nabla\bar u+\nabla\bar u^T\right)=\left(\ba u^1_x \qquad \frac{1}{2}(u^1_y+u^2_x) \\
    \frac{1}{2}(u^1_y+u^2_x) \qquad u^2_y \ea \right).\]

\begin{remark}
In the case of anisotropic materials, the matrix $\tilde \tau _t$
has a more complicated structure which can be found
(e.g., in \cite{loret-simoes-17}).
\end{remark}

  In order to obtain a closed system of PDEs for finding unknown functions, one needs to specify the fluxes across the PEM.

  In~the fluid phase, we have the following flux:
    \[{\bar j_{VF}} = {\theta _F}\frac{{\partial \bar u}}{{\partial t}}- k \nabla \cdot p^*
    \equiv {\theta _F}\frac{{\partial \bar u}}{{\partial t}}- k\left({\nabla  p - \sigma_1 \nabla c
    }\right).\]

  Here,  the first term means  the deformation contribution,  and the second
       is the volumetric fluid flux relative to the matrix calculated according to the extended Darcy's
       law with hydraulic conductivity  $k$.

     The volumetric  flux in the solid phase (matrix) is as follows:
    ${\bar j_{VM}} = {\theta _M}\frac{\partial \bar u}{\partial
    t}.$

In contrast to the fluxes defined above, the~solute flux through the
PEM includes the diffusive term reflecting Fick's law,  and  the
convective term as follows: \[{\bar j_S} = -D\nabla c +Sc \bar j_{VF}, \]
i.e., \[\bar j_S = - D\nabla c - Sk c (\nabla  p - \sigma_1 \nabla
c)+{\theta _F}c\frac{{\partial \bar u}}{\partial t},\] where
 $D$ is the  solute diffusivity, and
 $S<1$  is the so-called sieving coefficient of the solute in the PEM.
  In the case of homogeneous membranes, which can be considered as PEM,
  the coefficient $S = 1- \sigma$, where  $\sigma <1$  is the  reflection coefficient
  (see, e.g., \cite{che-sta-wa-14}). The latter is related to the osmotic pressure
  and it can be  identified that  $ \sigma_1 =\sigma RT$,
 where $R$ is the gas constant and $T$ is a given temperature.

Finally, if you go back to the  continuity equations  (\ref{2-1*})
and (\ref{2-3*}), the~fluxes
${\bar j_V}$  and ${\bar j_\rho }$  should  be related to the above defined fluxes as follows:
    \[{\bar j_V} = {\bar j_{VF}} + {\bar j_{VM}} = - k( {\nabla  p - \sigma_1 \nabla c }) + \frac{{\partial u}}{{\partial
    t}},\]
    \[{\bar j_\rho } = {\rho _F}{\bar j_{VF}} + {\rho _M}{\bar j_{VM}} = -k{\rho _F}( {\nabla  p - \sigma_1 \nabla c }) +
    \rho \frac{{\partial u}}{{\partial t}}.\]

Now, one realizes that equations (\ref{2-1*}) and (\ref{2-3*}) are algebraic consequences of equations (\ref{2-4})--(\ref{2-7}) written down for each phase of PEM;  therefore, those can be skipped.

 Thus, by inserting  the fluxes defined above into the governing  equations (\ref{2-4})--(\ref{2-10}) and making the relevant transformations and calculations, one arrives at    a  system of $n+4$
    PDEs for  unknown functions  $\bar u=(u_1,\dots,u_n)$, \ ${\theta _F}$, \ ${\rho}$, \  $c$, and $p$. Two other unknown functions, ${\rho _M}$ and ${\theta _M}$, are defined by the algebraic equations (\ref{2-8}).

  In the case of the 1D approximation ($n=1$),  the system obtained was studied in \cite{ch-wa-2020,ch-st-wa-24}.
 The rest of this study is devoted to the case when  $n=2$.

 \section{The  2D model and its  Lie symmetry  } \label{sec:3}

In the 2D space approximation,  the governing equations
of the model possess the following  form:
  \begin{equation}\label{4-7}
2\nabla \cdot  \bar u_t=k\Delta p^*,\ee \be \label{4-8} \rho
u^1_{tt}+u^1_t(\rho_t + \rho \nabla \cdot  \bar
u_t)=\lambda^*u^1_{xx} +\mu u^1_{yy}+(\lambda^*-\mu)
u^2_{xy}-p^*_{x} ,\ee \be \label{4-9} \rho u^2_{tt}+u^2_t(\rho_t +
\rho \nabla \cdot  \bar u_t)=\lambda^*u^2_{yy} +\mu
u^2_{xx}+(\lambda^*-\mu) u^1_{xy}- p^*_{y},\ee \be
\label{4-10}\rho_t+\bar u_t\nabla\rho=k(\rho^0_F - \rho)\Delta
p^*,\ee \be \label{4-11}{\theta_F}_t+\bar
u_t\nabla{\theta_F}=k(1-\theta_F)\Delta p^*,\ee \be
\label{4-12}\lf(c\theta_F\rg)_t+\bar u_t\cdot\nabla(c\theta_F)= D\Delta
c+ k(S-\theta_F)c\Delta p^*+ kS\nabla c\cdot\nabla p^*,
\end{equation}
where
 $p^*= p - \sigma_1c$, \ $\lambda^*=\lambda+2\mu$, $\Delta$ is
the Laplace operator, $\nabla \cdot \bar{f}\equiv f^1_{x}+f^2_{y}$, and the lower subscripts $t$ and $x$
denote differentiation with respect to these~variables.
Natural restrictions on the parameters arising above are as follows:
 \be\label{4-13}k> 0, \ D> 0, \ \rho^0_F > 0, \ 0 < S < 1, \ \lambda > 0, \ \mu > 0.
 \ee

\begin{remark}
Generally speaking, some limiting cases may occur when  some
parameters vanish. For example, $D\simeq 0$   when very large
molecules (e.g., albumin)  are dissolved in the fluid.
Here, we do not consider such limiting
cases.
\end{remark}

We remind the reader that the above equations were derived under the assumption that the PEM is an isotropic material. If this assumption is removed, then the Terzaghi effective stress tensor
(\ref{2-11}) is  more complicated and reads as follows:
\be \label{2-1}{\tilde \tau _{ter}} = -p^*I +E\epsilon,
 \end{equation}
  with  the symmetric matrix \cite{loret-simoes-17}
\begin{equation}\label{2-1b}
E=\left(\ba E_{11} \qquad E_{12} \qquad \sqrt{2}E_{13}\\ E_{12}
\qquad E_{22} \qquad \sqrt{2}E_{23}\\  \sqrt{2}E_{13} \
\sqrt{2}E_{23} \qquad 2E_{33} \ea \right)\ee and the strain tensor
\begin{equation}\label{2-1d}
 \epsilon=\frac{1}{2}\left(\nabla\bar u+\nabla\bar u^T\right)\Leftrightarrow
 \left(u^1_x,u^2_y, \frac{\sqrt{2}}{2}(u^1_y+u^2_x)\right)^T.\ee
 Here,   $E_{ii}>0, \
E_{ij}\geq0, i\neq j$.

\begin{remark}
In the case of isotropic materials,  the
above matrix essentially simplifies because
\begin{equation}\label{2-1c} E_{13}=E_{23}=0, \ E_{33}=\mu, \
E_{11}=E_{22}=\lambda + 2\mu, \  E_{12}=E_{22}-2E_{33}=\lambda, \end{equation}
where
  $\lambda$ and $\mu$ are the Lame coefficients.
\end{remark}

Thus, making a similar routine as the aforementioned  isotropic case,  the
governing equations for the fluid and solute transport in the PEM
can be derived in the following anisotropic case:
\begin{equation}\ba \label{2-2}
u^1_{tx}+u^2_{ty}=\frac{k}{2}\Delta p^*,\medskip \\
 (\rho_t+ \rho(u^1_{tx}+u^2_{ty}) )u^1_t
+\rho u^1_{tt}= \medskip \\
\hskip1cm -p^*_{x}+E_{11}u^1_{xx}+E_{33}u^1_{yy}+
E_{13}u^2_{xx}+E_{23}u^2_{yy}+2E_{13}u^1_{xy}+(E_{12}+E_{33})u^2_{xy},\medskip\\
 (\rho_t+ \rho(u^1_{tx}+u^2_{ty}) )u^2_t
+\rho u^2_{tt}= \medskip \\
\hskip1cm -p^*_{y}+E_{22}u^2_{yy}+E_{33}u^2_{xx}+
E_{13}u^1_{xx}+E_{23}u^1_{yy}+2E_{23}u^2_{xy}+(E_{12}+E_{33})u^1_{xy}, \medskip\\
\rho_t+\rho_xu^1_t+\rho_yu^2_t=k(\rho^0_F - \rho)\Delta p^*\\
{\theta_F}_t+{\theta_F}_xu^1_t+{\theta_F}_yu^2_t=k\lf(1-\theta_F\rg)\Delta p^*,\medskip\\
\lf(c\theta_F\rg)_t+\lf(c\theta_F\rg)_xu^1_t+\lf(c\theta_F\rg)_yu^2_t=D\Delta
c +k(S-\theta_F)c\Delta p^*+kS\nabla c\cdot\nabla p^*.\ea
\end{equation}
In the case of isotropic PEM, the governing equation (\ref{2-2})
coincide with  the PDE systems (\ref{4-7})--(\ref{4-12}).

\begin{theorem}\label{th-1} The nonlinear PDE system (\ref{2-2}) is invariant
w.r.t.  an infinitely-dimensional Lie algebra. By assuming  that all
parameters are arbitrary, the Lie algebra is
 generated by the  Lie symmetry operators as follows:
 \be\label{3-1*} \ba  \frac{\partial  }{\partial t},  \ \frac{\partial  }{\partial x},  \ \frac{\partial  }{\partial y}, \  c\frac{\partial  }{\partial c}, \medskip \\
g(t) \frac{\partial}{\partial p^*},\ G^1(x,y)\frac{\partial  }{\partial u^1}
+G^2(x,y) \frac{\partial  }{\partial u^2}.\ea \ee
 Here, $g(t)$ is
an arbitrary function, while the  functions $G^i, \ i=1,2$ form an
arbitrary  solution of the linear PDE system  as follows:  \[\ba 
E_{11}G^1_{xx}+E_{33}G^1_{yy}+
E_{13}G^2_{xx}+E_{23}G^2_{yy}+2E_{13}G^1_{xy}+(E_{12}+E_{33})G^2_{xy}=0,\medskip
\\ E_{22}G^2_{yy}+E_{33}G^2_{xx}+
E_{13}G^1_{xx}+E_{23}G^1_{yy}+2E_{23}G^2_{xy}+(E_{12}+E_{33})G^1_{xy}=0.
\ea\] \end{theorem} Depending on the parameters $E_{ij}$, there is
only a single extension of the principal  algebra (\ref{3-1*}). Thus, the extension only occurs  under the conditions
(\ref{2-1c}) (i.e., in the case
 of isotropic PEM).

 \begin{theorem}\label{th-2} The nonlinear PDE system (\ref{4-7})--(\ref{4-12}) is invariant w.r.t.
 an
infinitely-dimensional Lie algebra. By assuming  that all parameters are
arbitrary, the Lie algebra is
 generated by the  Lie symmetry operators as follows:
 \[ \ba  \frac{\partial  }{\partial t},  \ \frac{\partial  }{\partial x},  \ \frac{\partial  }{\partial y}, \  c\frac{\partial  }{\partial c}, \medskip \\
g(t) \frac{\partial}{\partial p^*},\ G^1(x,y)\frac{\partial
}{\partial u^1} +G^2(x,y) \frac{\partial  }{\partial u^2}, \medskip
\\ y\frac{\partial  }{\partial x}-x\frac{\partial  }{\partial
y}+u^2\frac{\partial  }{\partial u^1}-u^1\frac{\partial  }{\partial
u^2}.\ea \] Here,
the  functions $G^i, \ i=1,2$ form an arbitrary  solution of the
linear PDE system as follows:  \be\label{3-4}\ba  \lambda^*G^1_{xx}+\mu
G^1_{yy}+ (\lambda^*-\mu )G^2_{xy}=0,\medskip
\\
\mu  G^2_{xx}+\lambda^* G^2_{yy}+ (\lambda^*-\mu )G^1_{xy}=0. \ea\ee
\end{theorem} \textbf{Sketch of the proof of Theorems~\ref{th-1}
and~\ref{th-2}}. The proof is  based on the Lie infinitesimal
invariance criterion (see, e.g., Section 1.2.5 \cite{bl-anco-10}).
Since the PDE system (\ref{2-2}) only contains  constant parameters (i.e., one does not involve arbitrary functions),  the problem of Lie
symmetry classification is a mostly  technical routine. It mainly
involves  cumbersome calculations due to the detailed examination of
the determining equations under  restrictions (\ref{4-13}) and
$E_{ii}>0, \ E_{ij}\geq0, i\neq j$. The process requires careful
attention to the restrictions and the step-by-step calculation of
all possible symmetries.  This process  is omitted here;  however, we refer interested readers to the recent study \cite{ch-da-vo-24},
in which  nonlinear PDE system (\ref{4-7})--(\ref{4-12}) in 1D space case is studied and the relevant details for the Lie symmetry search are presented.
Notably, the computer algebra package Maple was unable to calculate
Lie symmetries of   PDE systems (\ref{2-2})  and
(\ref{4-7})--(\ref{4-12}).

\begin{remark}
The PDE system (\ref{3-4}) is reducible to the system of the
first-order PDEs \be\label{3-7}\ba  \lambda^*F^1_{x}-\mu
F^2_{y}=0,\medskip
\\
\lambda^* F^1_{y}+\mu F^2_{x}=0, \ea\ee by the substitution
\[ F^1(x,y) = G^1_x + G^2_y, \ F^2(t,x) = G^2_x -
G^1_y.\] System (\ref{3-7}) has the following  solution:
\[F^1=\mu\left(f_1(x+iy)+if_2(x-iy)\right), \
F^2=-\lambda^*\left(if_1(x+iy)+f_2(x-iy)\right),\] where $f_1$ and
$f_2$ are arbitrary functions, and $i=\sqrt{-1}.$  Notably, (\ref{3-7})
is  nothing else but   the well-known  Cauchy--Riemann system in a
non-standard form.
\end{remark}

According to the Lie continuous groups theory,  the  Lie symmetry operator
  \be\label{3-3}
 y\frac{\partial  }{\partial x}-x\frac{\partial  }{\partial y}+u^2\frac{\partial  }{\partial u^1}
 -u^1\frac{\partial  }{\partial u^2} \ee
 generates the one-parameter  Lie group
 \[
  (x',y')^T= A(x,y)^T,  \quad  ((u^1)',(u^2)')^T= A(u^1,u^2)^T, \]
  where $ A= \left(\ba  \cos\phi \quad -\sin\phi \\ \sin\phi \quad \cos\phi  \ea \right) $ is a rotation matrix, and $\phi$ is a group parameter. Thus, the above Lie symmetry reflects invariance of the system w.r.t.
 rotations.
   Interestingly, the Lie symmetry  (\ref{3-3}) has the same form as
 for the classical Navier--Stokes equations (in 2D approximation) \cite{Lloyd-1981}  and a tumor growth  model that was studied in
  \cite{ch-da-2020}.

\section{ The radially-symmetric case } \label{sec:4}

Here, we are looking for  solutions with radial symmetry of the
governing equations (\ref{4-7})--(\ref{4-12}), which are predicted by the
Lie symmetry (\ref{3-3}).  Thus,  one can rewrite the system in
polar coordinates, using the following transformations:
 \be\label{5-1ad}\begin{array}{l} x=r\cos\phi, \  y=r\sin\phi, \medskip \\
 u^1=w^1\cos\phi-w^{2}\sin\phi,\ u^2=w^1\sin\phi+w^{2}\cos\phi,\medskip \\
 p=P(t,r,\phi), \ \rho=\varrho(t,r,\phi), \ \theta_F=\Theta(t,r,\phi), \
 c=C(t,r,\phi),
 \end{array}\ee
which are typically used for the displacement vector $(u^1,u^2)$
(see, e.g.,  \cite{loret-simoes-17}).

 By applying transformation
(\ref{5-1ad}) to the  governing equations (\ref{4-7})--(\ref{4-12}), we
arrive at the  following system:
 \be\label{5-4ad} \ba 
\left(\left(rw^1\right)_r+w^{2}_{\phi}\right)_t=\frac{k}{2r}\,P_{\phi\phi}+\frac{k}{2}\,
\left(rP_r\right)_r,\medskip\\
w^1_{t}\left( \varrho_t+ \varrho w^1_{tr}+
 \frac{1}{r}\varrho w^1_{t}+\frac{1}{r}\varrho w^2_{t\phi}\right)
+\varrho
w^1_{tt}=-P_{r}+\lambda^*w^1_{rr}+\frac{\lambda^*}{r}\,w^1_{r}-
\frac{\lambda^*}{r^2}\,w^1+\medskip\\
\hskip2cm \frac{1}{r^2}\left((\lambda^*-\mu)rw^2_{r\phi}+\mu w^1_{\phi\phi}-(\lambda^*+\mu)w^2_{\phi}\right),\medskip\\
w^2_{t}\left(\varrho_t+\varrho w^1_{tr}+
 \frac{1}{r}\varrho w^1_{t}+\frac{1}{r}\varrho w^2_{t\phi}\right)
+\varrho w^2_{tt}=-\frac{1}{r}P_{\phi}+\mu
w^2_{rr}+\frac{\mu}{r}\,w^2_{r}-
\frac{\mu}{r^2}\,w^2+\medskip\\
\hskip2cm \frac{1}{r^2}\left((\lambda^*-\mu)rw^1_{r\phi}+\lambda^* w^2_{\phi\phi}+(\lambda^*+\mu)w^1_{\phi}\right),\medskip\\
 r\varrho_t+r\varrho_rw^1_t+\varrho_\phi
w^2_t=k(\rho^0_F -
\varrho)\left(\frac{P_{\phi\phi}}{r}+\left(rP_r\right)_r\right),\medskip\\
r\Theta_t+r\Theta_rw^1_t+\Theta_\phi w^2_t=k(1 -
\Theta)\left(\frac{P_{\phi\phi}}{r}+\left(rP_r\right)_r\right),\medskip\\
r\left(C\Theta\right)_t+r\left(C\Theta\right)_rw^1_t+\left(C\Theta\right)_\phi
w^2_t=\medskip\\
\hskip2cm D\left(\frac{C_{\phi\phi}}{r}+\left(rC_r\right)_r\right)+
k(S-\Theta)C\left(\frac{P_{\phi\phi}}{r}+\left(rP_r\right)_r\right)+\frac{kS}{r}\left(C_\phi
P_\phi+r^2C_rP_r\right).
 \ea\ee

Simultaneously, the Lie symmetry (\ref{3-3})   essentially  takes the
simpler form \be\label{5-2ad} J=-\p_{\phi}.\ee The relevant ansatz
generated by operator (\ref{5-2ad}) leads to exact  solutions
with radial symmetry 
and has the form
\be\label{5-3ad}\ba  w^1=w^1(t,r), \ w^2=w^2(t,r), \ P=P(t,r),\\
\varrho=\varrho(t,r),  \ \Theta=\Theta(t,r), \ C=C(t,r). \ea\ee
By substituting  ansatz  (\ref{5-3ad}) into (\ref{5-4ad}), one obtains
the following system of $(1+1)$-dimensional equations:
 \be\label{5-4ad0} \ba 
\left(rw^1\right)_{rt} = \frac{k}{2}\,
\left(rP_r\right)_r,\medskip\\
w^1_{t}\left( \varrho_t+ \varrho w^1_{tr}+
 \frac{1}{r}\varrho w^1_{t}\right)
+\varrho
w^1_{tt}=-P_{r}+\lambda^*w^1_{rr}+\frac{\lambda^*}{r}\,w^1_{r}-
\frac{\lambda^*}{r^2}\,w^1 \medskip,\\
w^2_{t}\left(\varrho_t+\varrho w^1_{tr}+
 \frac{1}{r}\varrho w^1_{t}\right)
+\varrho w^2_{tt}=\mu w^2_{rr}+\frac{\mu}{r}\,w^2_{r}-
\frac{\mu}{r^2}\,w^2, \medskip\\
 r\varrho_t+r\varrho_rw^1_t=k(\rho^0_F -
\varrho)\left(rP_r\right)_r,\medskip\\
r\Theta_t+r\Theta_rw^1_t=k(1 -
\Theta)\left(rP_r\right)_r,\medskip\\
r\left(C\Theta\right)_t+r\left(C\Theta\right)_rw^1_t=
D\left(rC_r\right)_r+ k(S-\Theta)C\left(rP_r\right)_r+kS\,rC_rP_r.
 \ea\ee

\textbf{Example.} Let us consider the PEM in the form of a circle (ring)
with a radius $R_0$ (see Figure~\ref{fig:2}). We assume that all pores
in the circle are connected by capillaries/fibers (this is not shown
on the picture), saturated by water, and the system is in
steady-state (the gravity force is not taken into account). At the
time moment $t=0$, a load $F_0$ begins to shrink (squeeze) the
circle in the radially-symmetric direction. As a result, the
boundary of the circle starts to move and simultaneously one expects
a water outflow throughout the moving boundary. In the porous
circle, the diffusion process of the water towards the boundary
takes place. At the time moment $t=T$, the load stops to act.
Finally, one obtains a new porous circle with a radius $r_{st}<R_0$.
In the  new steady-state, the porosity of the circle will be smaller
because we assume  a constant pressure $p_a$ at the ring surface all
time; however, the matrix volume (actually it is the square) is
still the same as at the moment $t=0$.

 \begin{figure}
    \centering
    \centerline{\includegraphics[width=5cm]{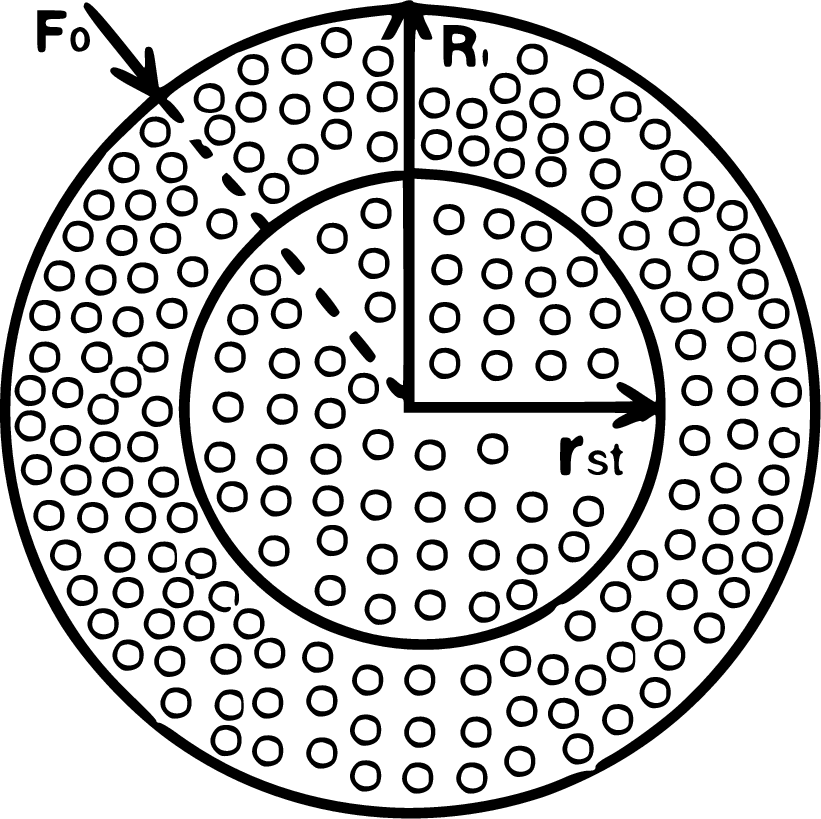} }\vspace{0.5cm}
    \caption[A]{PEM in the form of a circle (ring)
with the initial  radius $R_1$ and the radius $r_{st}$ after
deformation caused by the load $F_0$. }
    \label{fig:2}
\end{figure}
 \begin{figure}
    \centering
    \centerline{\includegraphics[width=5cm]{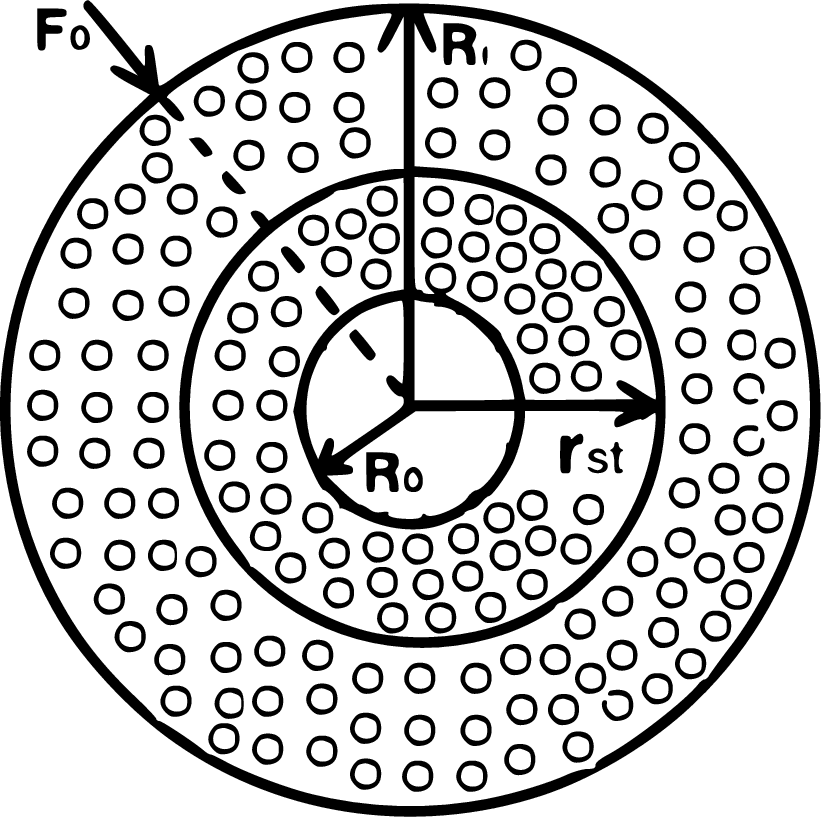} }\vspace{0.5cm}
    \caption[A]{PEM in the form of an annulus
with the initial outer  radius $R_1$ and the radius $r_{st}$ after
deformation caused by the load $F_0$. }
    \label{fig:1}
\end{figure}

The process described above allow us to simplify the governing equation
(\ref{5-4ad}). In fact,  deformation will  only take place in the
radially-symmetric direction, while the tangential deformation is
zero (or negligibly small); therefore, one may set $w^2=0$
(i.e., the
third equation in (\ref{5-4ad0}) vanishes). Moreover, we consider
pure water (i.e., there is no solute transport); therefore, the last
equation can be skipped because one may set $C=0$. As a result, the
mathematical model that describes  the deformation of the PEM in the form of
the  circle consists of the following governing equations:
\begin{equation}\ba \label{5-5ad}
w_{tr}+\frac{1}{r}w_{t}=\frac{k}{2}\left(P_{rr}+\frac{1}{r}\,P_{r}\right),\medskip \\
 w_{t}\,\Big( \varrho_t+ \varrho w_{tr}+
 \frac{1}{r}\varrho w_{t}\Big)
+\varrho w_{tt}=
 -P_{r}+\lambda^*w_{rr}+\frac{\lambda^*}{r}\,w_{r}-
\frac{\lambda^*}{r^2}w,\medskip\\
 \varrho_t+\varrho_r w_{t}=k(\rho^0_F - \varrho)\left(P_{rr}+\frac{1}{r}\,P_{r}\right),\medskip\\
\Theta_t+\Theta_r
w_{t}=k\lf(1-\Theta\rg)\left(P_{rr}+\frac{1}{r}\,P_{r}\right),\ea
\end{equation}
where the notation $w=w^1$ is used.


In order to complete
the mathematical model, one needs to specify   boundary conditions
and initial profiles.

The boundary conditions are as follows:
\begin{equation}\ba \label{5-1}
r=0: \quad  \frac{\p P}{\p r}=0, \ \frac{\p \varrho}{\p r}=0, \
\frac{\p \Theta}{\p r}=0, \ w =0,\\
\medskip\\
r=S(t): \quad  P=p_a, \  w =S(t)-R_0, \ea \end{equation} where
$S(t)$ is an unknown function that essentially depends  on the load
$F_0$ and has the specified values  $S(0)=R_0$  and $S(T)=r_{st}.$

The initial conditions are as follows:
\begin{equation}\label{5-2}
t=0: \quad  P=p_a, \ \varrho= \varrho_0(r), \ \Theta=\Theta_0(r), \
w =0,\end{equation} and the condition for the final displacement of
the surface is
\begin{equation}\label{5-3}
t=T: \quad   w =r_{st}-R_0,\end{equation}
 where $ \varrho_0(r)$ and $\Theta_0(r)<1$ are given
positive functions, and $R_0$ and $r_{st}$ are known positive constants.

In order to take the load
$F_0$ (generally speaking, it can be  a function of time and space)
acting on the moving boundary into account, we need to specify the Terzaghi
effective stress tensor in the radially-symmetric case. Taking  the assumption about radial symmetry and the strain
tensor matrix presented in \cite{taber-04} (see Appendix A) in the
cylindrical coordinates into
account, the matrix $\varepsilon$ (see (\ref{2-1d}))
is equivalent to the vector \[ \left(\frac{\p w}{\p r},\frac{ w}{
r}, 0 \right)^T\!. \] Thus, using formulae (\ref{2-1}), (\ref{2-1b}) and
(\ref{2-1c}), the Terzaghi effective stress tensor can be derived in
the form \[ \left(-P + \lambda^*\frac{\p w}{\p r}+\frac{ \lambda}{
r}w, -P + \lambda\frac{\p w}{\p r}+\frac{ \lambda^*}{ r}w, 0
\right)^T\!.
\] It means that  the tensor matrix $\tilde \tau _{ter}$ is a
diagonal matrix with the nonzero elements
 \[\tau_{11}= -P + \lambda^*\frac{\p w}{\p
r}+\frac{ \lambda}{ r}w, \quad \tau_{22}= -P + \lambda\frac{\p w}{\p
r}+\frac{ \lambda^*}{ r}w. \]

 Because the load $F_0$ only pushes the
boundary  along the radius $r$, one readily obtains the
condition $ \tilde \tau _{ter}\bar n =(\tau_{11},\tau_{22})\cdot (1,0)=-\frac{F_0}{2\pi r}$, where $\bar n =(1,0)$
is normal to the boundary. Thus,  we arrive at the following additional
boundary condition:
 \be \label{5-4} r=S(t): \quad \lambda^*\frac{\p
w}{\p r}+\frac{ \lambda}{ r}w = p_a -\frac{F_0}{2\pi r}.
 \end{equation}

Thus, we obtain the BVP with moving boundary that  consists of the
governing equations (\ref{5-5ad}), the boundary conditions (\ref{5-1}) and
(\ref{5-4}),  and the initial conditions (\ref{5-2})--(\ref{5-3}).

The above example can be generalized by replacing the circle by the
annulus (see Figure~\ref{fig:1}) as follows:
\[ \Omega= \{ (r,\phi): \ R_0< r< R_1; 0\leq \phi <2\pi \}. \]
In this case, we may assume the Dirichlet conditions on the interior boundary of the annulus as follows:
\begin{equation}\label{5-7}
r=R_0: \quad  P=p_a, \ \varrho= \varrho_0, \ \Theta=\Theta_0, \
w =0.\end{equation}
The last condition in (\ref{5-7}) means that the interior of the annulus is fixed
(i.e., zero displacement at $r=R_0$).

 \begin{remark}
 A simpler example for a water-saturated soil
 layer is studied in \cite{coussy-2010} (see Chapter 5). The bone
 deformation was studied in \cite{daria-et-al-2001} using the bone layer
 in the form of circle; however, the authors did not present the
 mathematical model in detail.
 \end{remark}

 \section{ Stationary exact solution } \label{sec:5}

Constructing the exact solution  of the nonlinear BVP that was
formulated above is a highly nontrivial problem.
 Here, we consider the stationary case  that can be treated  in a straightforward way.
 In the stationary case (i.e., all unknown functions do not depend on the time variable), the nonlinear  system (\ref{5-5ad}) reduces to two linear second-order ODEs with the general solution
 \[
 P = P_0 + C_0\ln r, \quad w = \frac{C_0}{2\lambda^*}r\ln r+C_1r+C_{-1}r^{-1},
 \]
 while $\varrho(r) $ and $\Theta(r)$ are arbitrary smooth functions ($P_0$ and $C$ with indices are arbitrary constants).

Let us  consider the above example for the PEM in the form of  the annulus $\Omega$.
 The stationary  solution satisfies the boundary conditions (\ref{5-7})
 if the following hold:
\be\nonumber\ba  p_a=P_0+C_0\ln R_0 \ \Leftrightarrow \
C_0=\frac{p_a-P_0}{\ln R_0}, \\
\frac{p_a-P_0}{2\lambda^*}\,R_0+C_1R_0+C_{-1}\,R_0^{-1}=0 \
\Leftrightarrow \
C_{-1}=\left(\frac{P_0-p_a}{2\lambda^*}-C_1\right)R_0^2.
 \ea\ee
Let us assume that  we have a steady-state (stationary)  position of the annulus with
the following  boundary conditions:
\be\label{5-8} r=r_{st}\,: \ P=p_{st}, \ w=r_{st}-R_1.\ee
Thus, using the above boundary conditions, we specify $P_0$ and $C_1$ as follows:
\[P_0+\frac{p_a-P_0}{\ln R_0}\,\ln r_{st}=p_{st} \ \Leftrightarrow \ P_0=\frac{p_{st}\ln R_0 -p_a \ln r_{st}}{\ln R_0- \ln r_{st}}, \  r_{st}\not=R_0.\]
In the case $r_{st}=R_0$,  the condition $p_{st}=p_a$ takes place.
From the  second condition in (\ref{5-8})
\[ \frac{p_a-P_0}{2\lambda^*\ln R_0}\,r_{st}\ln
 r_{st} - \frac{(P_0-p_a)R_0^2}{2\lambda^* r_{st}}+C_1r_{st}+\left(\frac{P_0-p_a}{2\lambda^*}-C_1\right)\frac{R_0^2}{r_{st}}=r_{st}-R_1,\]
 one  defines
 \[C_1=\left(\frac{P_0-p_a}{2\lambda^* \ln R_0}\,r_{st}\ln
 r_{st}- \frac{P_0-p_a}{2\lambda^*}\frac{R_0^2}{r_{st}}+r_{st}-R_1\right)\frac{r_{st}}{r_{st}^2-R_0^2},\]  where $r_{st}\neq R_0$.

 If $r_{st}=R_0$, then
 $r_{st}=R_1$, and
 $C_1$ is an arbitrary constant; therefore,  we arrive at an unrealistic situation where   $r_{st}=R_1=R_0$ (i.e., the annulus degenerates into a circle).

 Thus,
\be\label{5-6}\ba P= P_0 +(p_a-P_0)\frac{\ln r}{\ln R_0},\medskip\\
w=\frac{p_a-P_0}{2\lambda^*}\,r\frac{\ln r}{\ln
R_0}+C_1r+\left(\frac{P_0-p_a}{2\lambda^*}-C_1\right)\frac{R_0^2}{r},\ea\ee
 where \be\label{5-9}P_0= \frac{p_{st}\ln R_0 -p_a \ln r_{st}}{\ln R_0- \ln r_{st}}=p_a+\frac{p_{st}-p_a}{1-\frac{\ln r_{st}}{\ln
 R_0}},\ee
 \[C_1=\left(\frac{p_{st}-p_a}{2\lambda^*(\ln R_0-\ln r_{st})}\,\left(r_{st}\ln r_{st}- \frac{R_0^2\ln R_0}{r_{st}}\right)+r_{st}-R_1\right)\frac{r_{st}}{r_{st}^2-R_0^2}.\]
By substituting  (\ref{5-9}) into (\ref{5-6}), we arrive at the following:
\be\label{5-6*}\ba   P=\frac{p_{st}\ln R_0-p_a\ln r_{st}}{\ln R_0-\ln r_{st}}-\frac{p_{st}-p_a}{\ln R_0-\ln r_{st}}\ln r,\\
 w=\frac{1}{2\lambda^*}\frac{p_{st}-p_a}{\ln R_0-\ln r_{st}}\left(\frac{R_0^2}{r}\ln R_0-r\ln r\right)+C_1\left(r-\frac{R_0^2}{r}\right),\ea\ee
 where $r_{st}\neq R_0.$

 Finally, we need to determine the parameter $r_{st}$  that is an analogue of the unknown function $S(t)$ in the nonstationary case.  This can be done using a modification of the boundary condition  (\ref{5-4}) in the following form:
\be  \label{5-10} r=r_{st}: \quad \lambda^*\frac{\p w}{\p r}+\frac{
\lambda}{ r}w = p_a -\frac{F_0(r)}{2\pi r}.
 \end{equation}
 By substituting  the function $w$ from  (\ref{5-6*}) into  (\ref{5-10}), one obtains a  cumbersome   transcendent  equation to find  $r_{st}$ that is omitted here.

A simpler situation occurs if one applies the boundary conditions that involve
the Neumann condition (zero flux) for the pressure $P$:
\[r=R_0: \ \frac{\p P}{\p r}=0, \
w=0.\]
The above conditions immediately produce the following:
\[\frac{\p P}{\p r}\Bigg|_{r=R_0}=\frac{C_0}{R_0}=0 \ \Rightarrow \ C_0=0,\]
\[w\big|_{r=R_0}=C_1R_0+C_{-1}R_0^{-1}=0 \ \Rightarrow \ C_{-1}=-C_1R_0^2.\]
Moreover, the boundary conditions  (\ref{5-8}) lead to the following:
\[P_0=p_{st}, \ C_1\left(r_{st}-R_0^2r_{st}^{-1}\right)=r_{st}-R_1.\]
Therefore,  we arrive at the following exact solution: \be
\label{5-11} P=p_{st}, \
w=\frac{r_{st}\left(r_{st}-R_1\right)}{r_{st}^2-R_0^2}\left(r-\frac{R_0^2}{r}\right),
\quad  r_{st}\neq R_0. \ee In order determine the parameter
$r_{st}$, we use the boundary condition    (\ref{5-10}). In the case
of the above function $w$,  condition (\ref{5-10})  reduces to the
following algebraic equation:
\[ 2(\lambda+\mu) \frac{r_{st}\left(r_{st}-R_1\right)}{r_{st}^2-R_0^2}
 +2\mu \frac{R_0^2\left(r_{st}-R_1\right)}{r_{st}(r_{st}^2-R_0^2)}
 =p_a-\frac{F_0(r_{st})}{2\pi r_{st}}. \]
In the simplest case, namely $F_0=const$, the above equation reduces
to the following cubic polynomial: \be \label{5-12}
P_3(r_{st})\equiv \left(\lambda+\mu- \frac{p_a}{2}\right)r_{st}^3+
\left(\frac{F_0}{4\pi} - (\lambda+\mu)R_1\right)r_{st}^2
+\left(\mu+\frac{p_a}{2}\right) R_0^2r_{st}- \left(\frac{F_0}{4\pi}
+\mu R_1\right)R_0^2=0.
 \ee
 Therefore, depending on the coefficients, the above equation has one, two,  or three real roots.
  Notably, $r_{st}=R_1$ if $F_0=2\pi R_1p_a$.

 For simplicity, we set  $p_a=0$ in  what follows. Obviously, assuming $r_{st}=R_1$, one
  obtains $F_0=0$ (i.e., there is no deformation because  nothing is acting on the annulus).
 The simplest nontrivial case occurs when $ F_0 =4\pi (\lambda+\mu)R_1$. In this case, it
 can be easily proven  that there exists a unique positive
  root $r_{st}=r_*$  that belongs  to the interval $( R_0, R_1)$.
 Thus, one may claim that the annulus was shrinking before the steady-state arrived.

 If  $ F_0 >4\pi (\lambda+\mu)R_1$, then $r_* \in ( R_0, R_1)$ as
 well. Let us prove this.
The left-hand-side  of (\ref{5-12}) is the  third-order polynomial $P_3(r_{st})$
  with critical points that satisfy
 the following quadratic equation:
 \[3(\lambda+\mu)r_{st}^2+ 2\kappa r_{st} +\mu R_0^2=0,
  \]
where  $ \kappa =\frac{F_0}{4\pi} - (\lambda+\mu)R_1 >0$. By solving
the above equation,  one obtains the following:
\[{r_{st}}_{+,-}= \frac{1}{3(\lambda+\mu)}\left(- \kappa \pm
\sqrt{\kappa^2 - 3(\lambda+\mu)\mu R_0^2}\right). \] Now, one realizes
that real parts of both critical points are negative.
 The point ${r_{st}}_{-}$ is the  point of a local maximum  and  ${r_{st}}_{+}$ is
 the  point of a local minimum provided $\kappa^2 - 3(\lambda+\mu)\mu R_0^2 > 0$. Thus,   the
polynomial is a strictly increasing function for all $ {r_{st}}>
{r_{st}}_{+}$. In the case
 of complex values of ${r_{st}}_{\pm}$, the
polynomial $P(r_{st})$ is a strictly increasing function on any
interval. In both cases, the polynomial must have a real root $r_*$.
Moreover, $r_*> {r_{st}}_{+}$ if ${r_{st}}_{+}$ is a real number and
$r_*$ is a unique real root if ${r_{st}}_{+}$ is complex.

 Finally, we simply calculate $P_3(R_0)=R_0^2(\lambda+2\mu)(R_0-R_1)
 < 0$  and $P_3(R_1)=\frac{F_0}{4\pi}(R_1^2 - R_0^2)>0$;
therefore,  $r_* \in ( R_0, R_1)$.
 Thus,  the annulus was shrinking before the steady-state arrived.

  \begin{remark} In the case when  $p_a>0$,  the situation is quite similar. By assuming  $ F_0 \geq 4\pi (\lambda+\mu)R_1$,
  it can be shown that    real parts of both critical points again are negative. In this case $P_3(R_0)=R_0^2(\lambda+2\mu)(R_0-R_1)
 < 0$  and $P_3(R_1)=\Big(\frac{F_0}{4\pi}-\frac{p_aR_1}{2}\Big)(R_1^2 - R_0^2)$, one needs the additional restriction $F_0> 2\pi p_aR_1$
 that guaranties that the annulus shrinking (i.e.,   $r_* \in ( R_0, R_1)$).
   \end{remark}

 \begin{figure}
    \centering
    \centerline{\includegraphics[width=8cm]{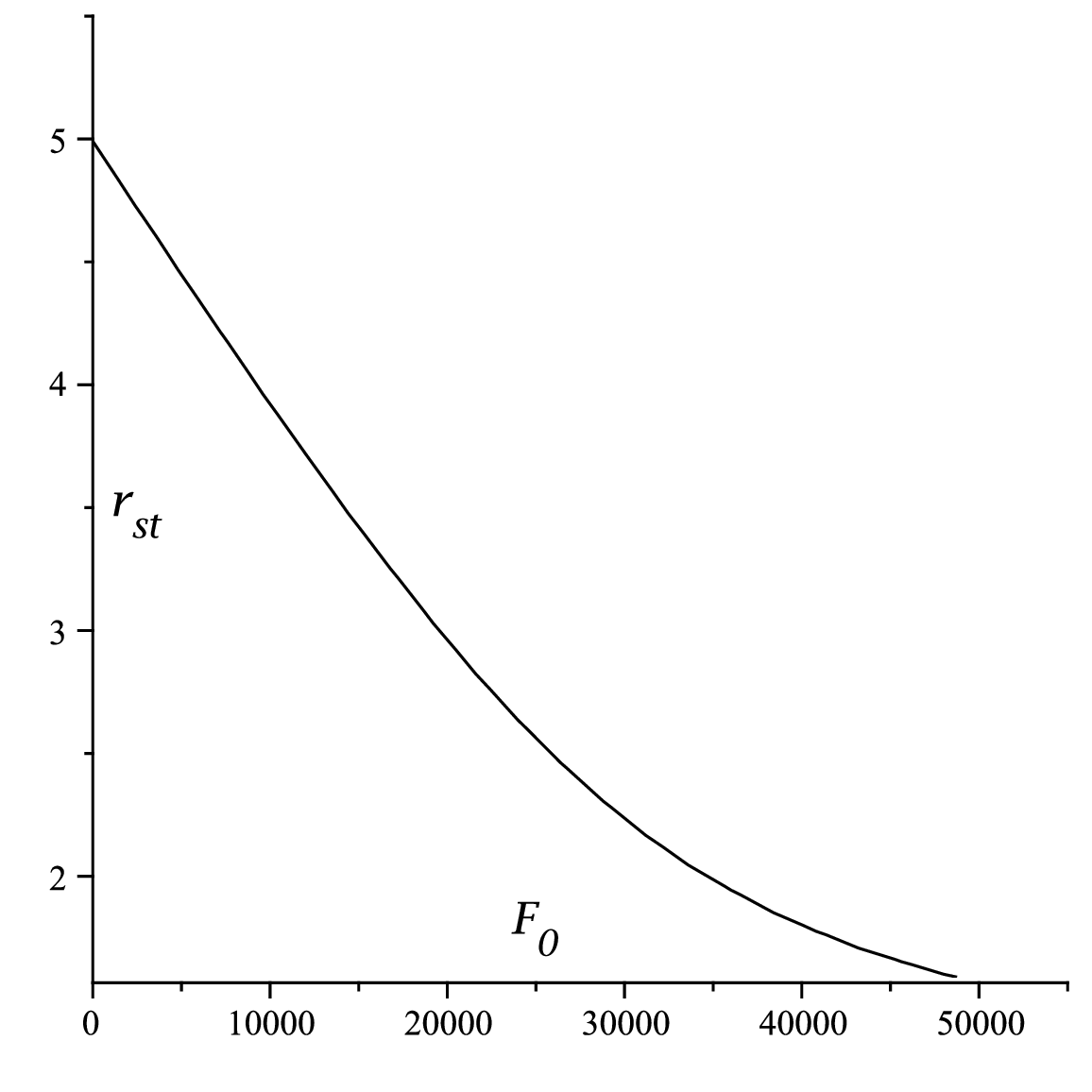} }\vspace{0.5cm}
    \caption[A]{The curve represents  the function  $r_{st}\left(F_0\right)$ (see Eq (\ref{5-12})
      for the parameters $ R_0=~1\ mm, \ R_1=5\ mm, \ \lambda=684\ mmHg, \ \mu=15 \ mHg).$}
    \label{fig:3}
\end{figure}

 Let us present example of calculations of
deformation of a tumor   in the form of the annulus $\Omega$ (i.e.,
assuming that necrosis already started; therefore, tumor  cells in the
domain $\{(r, \varphi): r< R_0 \}$ are dead). We specify the
parameters as follows: $ R_0= 1 \ mm, \ R_1= 5 \ mm, \ \lambda=684 \ mmHg,
\ \mu=15 \ mmHg.  $ Note that the Lame coefficients were taken from
\cite{netti-et-al-1997}, while the size of the tumor can
essentially vary  so that average  values were selected.

\begin{figure}
    \centering
    \centerline{\includegraphics[width=8cm]{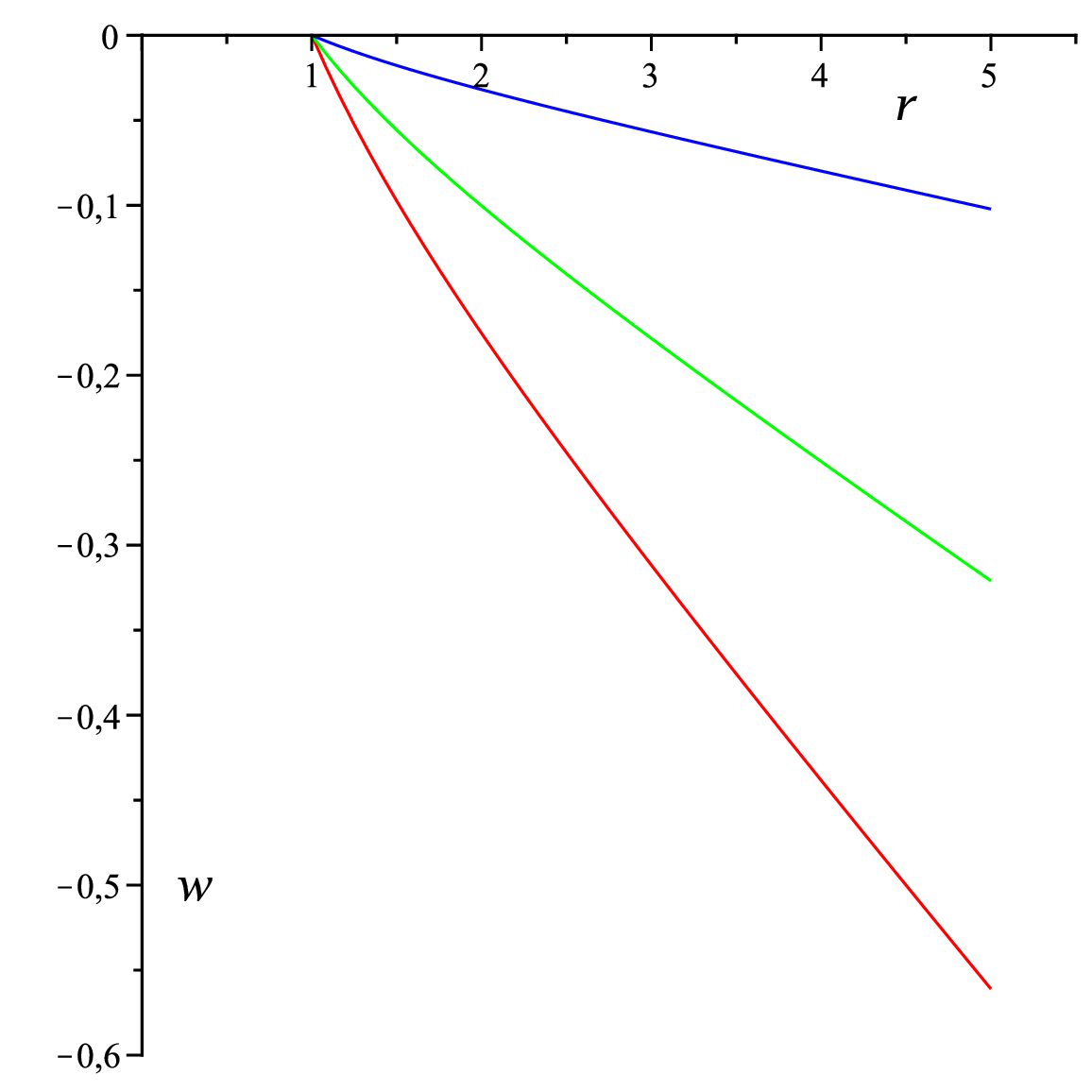} }\vspace{0.5cm}
    \caption[A]{The curves
represent deformations $w$  (see (\ref{5-11})) for $r_{st}=4.9 \ mm$
(blue curve), $r_{st}=4.7 \ mm$ (green curve), $r_{st}=4.5 \ mm$ (red
curve), and  $R_0=1 \ mm, \ R_1=5 \ mm$.}
    \label{fig:4}
\end{figure}

\begin{figure}
    \centering
    \centerline{\includegraphics[width=8cm]{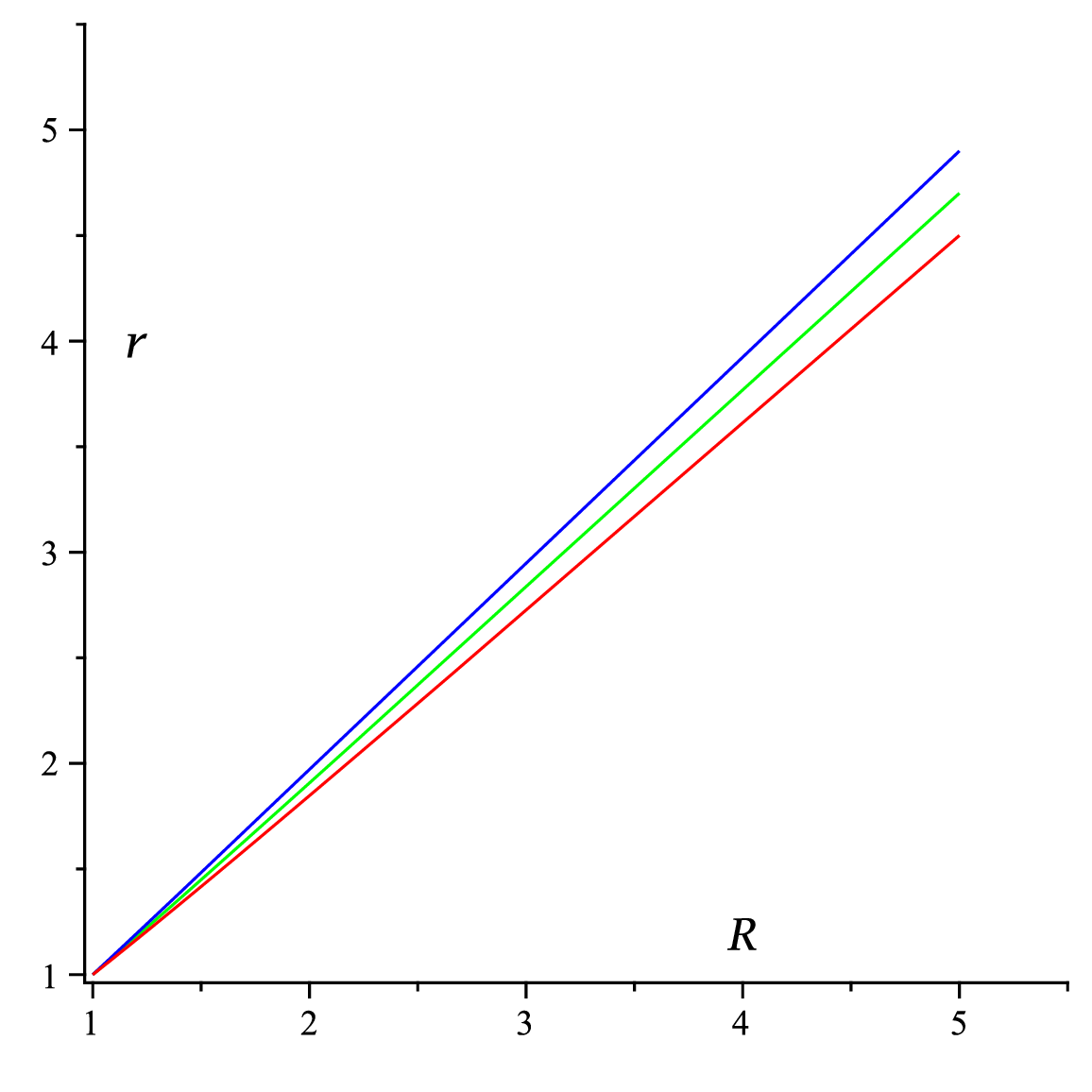} }\vspace{0.5cm}
    \caption[A]{The curves
represent the function $r(R)$ (see (\ref{5-14})) for $r_{st}=4.9 \ mm$
(blue curve), $r_{st}=4.7 \ mm$ (green curve), $r_{st}=4.5 \ mm$ (red
curve), and  $R_0=1 \ mm, \ R_1=5 \ mm$. }
    \label{fig:5}
\end{figure}

 In Figure~\ref{fig:3}, the curve represents  values of $r_{st}$
in the steady-state position of the annulus depending on the load
$F_0\geq0$. The curve was built  using the cubic  equation
(\ref{5-12}). In Figure~\ref{fig:4},  the curves represent
deformations as functions of $r$  for three  different values of
$r_{st}$. The second formula from (\ref{5-11}) was used  to draw
these plots. In Figure~\ref{fig:5}, the steady-state position of
each circle of the radius $r$ is presented as a function of the
initial steady-state position $R$. We note that the second formula
from (\ref{5-11}) can be rewritten in the following form (see
(\ref{2-2*})): \be \label{5-13} r-R
=\frac{r_{st}\left(r_{st}-R_1\right)}{r_{st}^2-R_0^2}\left(r-\frac{R_0^2}{r}\right).
\ee By solving  the algebraic equation (\ref{5-13}) with respect to
the $r$, one easily obtains two possibilities:
\[ r = \frac{R\pm\sqrt{R^2+4A(1+A)R_0^2}}{2(1+A)}. \]
Because $r$ must be positive, we fix the upper sign as follows:
 \be \label{5-14} r = \frac{R+\sqrt{R^2+4A(1+A)R_0^2}}{2(1+A)}, \quad A=\frac{r_{st}(R_1-r_{st})}{r_{st}^2-R_0^2}>0.
 \ee

\section{ Conclusions } \label{sec:6}

 A
mathematical model for PEM with the
variable volume was developed in a multidimensional case.  Governing
equations of the model were constructed using the continuity
equations, which reflect the well-known physical laws. The
deformation vector was specified using  the Terzaghi effective stress
tensor. As a result,  a model  based on   a  system of $n+4$
    partial differential equations for  unknown functions
     $\bar u=(u_1,\dots,u_n)$ (deformation vector), \ ${\theta _F}$ (porosity), \ ${\rho}$ (density), \  $c$ (concentration),
      and $p$ (hydrostatic pressure) was constructed.

 The model was studied in the $2D$ space
approximation by analytical methods. Using the classical Lie method,
it was proven  that the governing equations, which form a nonlinear
six-component system, admit highly nontrivial Lie symmetries leading
to an infinite-dimensional Lie algebra. It was shown that the
infinite-dimensional Lie algebra admits an extension in the case of
PEM  formed by isotropic material. Interestingly, the additional
Lie symmetry is the  same  as
 for the classical Navier--Stokes equations (in the $2D$ approximation)
 and  a model for tumor growth modeling \cite{Lloyd-1981, ch-da-2020}.

The radially-symmetric case was studied in detail. It was shown how
correct boundary conditions are constructed in the case of PEM in
the form of a ring that is shrinking (expanding). As a result, the
BVP  with a moving boundary that describes  the ring deformation  was
constructed; additionally, the  relevant BVP for the annulus deformation was
presented.  The nonlinear problems obtained were analytically
solved in the stationary case, using the standard method to solve
ODEs. In particular, the analytical formulae for unknown
deformations and an unknown radius of the annulus  were calculated.
It was shown how  deformation  and an unknown radius of PEM after
shrinking can  be exactly calculated by relevant formulae.
 Illustrative plots  were presented for the parameters,
which are  typical   for tumor  tissue deformation. Current research focuses on searching  for time-dependent solutions of the model
that admit plausible interpretations  such as  the stationary solutions
derived in this study. Symmetry-based methods \cite{bl-anco-10,
arrigo15, ch-se-pl-2018} will be applied in order to construct such
solutions.

As it is shown in this study, an infinite-dimensional Lie algebra
occurs  for   basic equations of the model. However, if one goes
to boundary conditions, then this wide invariance can be broken by
the geometry of the domain in question (see, a detailed discussion on this matter in
\cite{ch-ki-2015}). Therefore, one needs to identify domains (beyond ring
and annulus) in which the  nonlinear PDE system
(\ref{4-7})--(\ref{4-12}) still admits a nontrivial Lie symmetry.
For this, an  additional analysis is needed.  This lies beyond
the scope of this study and will be left for a future paper.

\section{Acknowledgments} R. Cherniha is grateful  to  Kostas Soldatos (University of
Nottingham) for useful comments.
The authors are grateful to unknown reviewers for some useful comments.
 R. Cherniha acknowledges that this
research was  partly funded by  the British Academy (Leverhulme
Researchers at Risk Research Support Grant LTRSF-100025).  V.D.
acknowledges that this research  was  supported by a grant from the
Simons Foundation (1290607, V.D.).

\end{document}